\documentclass[journal]{IEEEtran}
\usepackage{xcolor, soul}
\sethlcolor{yellow}
\usepackage{hyperref}
\usepackage{multirow}
\usepackage{balance}
\ifCLASSINFOpdf
\else
   \usepackage[dvips]{graphicx}
\fi
\usepackage{url}

\usepackage{graphicx}
\usepackage[space]{cite}

\def\BibTeX{{\rm B\kern-.05em{\sc i\kern-.025em b}\kern-.08em
    T\kern-.1667em\lower.7ex\hbox{E}\kern-.125emX}}

\begin{document}

\title{Low-Consumption Partial Transcoding by HEVC}

\author{Mohsen Abdoli, F\'{e}lix Henry, and Gordon Clare
\thanks{M. Abdoli, F. Henry and G. Clare are with IRT b$<>$com, Cesson-S\'{e}vign\'{e}, France. F. Henry and G. Clare are also with Orange Innovation, Cesson-S\'{e}vign\'{e}, France. (e-mail: firstname.lastname@b-com.com).}}

\markboth{IEEE Signal Processing Letters, VOL. X, NO. X, MONTH 2023}
{Shell \MakeLowercase{\textit{et al.}}: Bare Demo of IEEEtran.cls for IEEE Journals}
\maketitle

\begin{abstract}
 A transcoding scheme for the High Efficiency Video Coding (HEVC) is proposed that allows any partial frame modification to be followed by a partial re-compression of only the modified areas, while guaranteeing identical reconstruction of non-modified areas. To this end, first, syntax elements of all Coding Units (CU) in the frame are parsed and decoded according to their scan order. Then CUs that are collocated with a replaced area are re-encoded with new content to generate a partial set of new syntax elements. In order to avoid spatial propagation of the decoding mismatch due to the new content, CUs on the border of the replaced area are losslessly coded such that reconstruction of immediately neighboring CUs in the scan order are protected from the modification. The proposed method has been implemented on top of the HEVC test Model (HM) in All-Intra (AI) coding configuration and experiments show that, depending on the test parameters, it can offer both a bitrate saving (up to 4\% in terms of BD-BR) and a transcoding acceleration (up to 83\%) compared to a full transcoding scheme.
\end{abstract}


\IEEEpeerreviewmaketitle

\section{Introduction}

\IEEEPARstart{I}{n} various video communication systems, it is common for a single source content to undergo multiple encoding processes before reaching the end-user's display. This operation %, illustrated in Fig.~\ref{fig:full-transcode}, 
is commonly referred to as ``transcoding'' and can take place at various stages within the video delivery chain, depending on the purpose. Typically, transcoding is applied to an already encoded video stream to address capacity limitations of either the underlying network (\textit{e.g.}, bandwidth) or the receiving device (\textit{e.g.}, resolution) \cite{gao2021video}. A specific use case of transcoding, targetted in this paper, is where the coded bitstream is transcoded to replace a relatively small area of video frames in relation to the overall video resolution. Some examples of such use cases include: 1) Logo replacement/removal/enhancement, 2) Replacement of hardcoded closed captions (\textit{e.g.}, language change), 3) Personalized advertisement insertion and 4) Insertion of a sign language interpreter. In all these scenarios, the initial source content is encoded once at the sender's side and is modified slightly at later stages of the delivery chain.

% \IEEEPARstart{I}{n} different video communications systems, it often happens that a single source content is encoded more than once before reaching its end-user’s display. This operation, depicted in Fig.~\ref{fig:full-transcode}, is widely known as ``transcoding'' and can occur in different stages of a video delivery chain, depending on the intention. Most typically, transcoding is applied on an already coded video stream in order to cope with the capacity limitations of either underlying network (e.g. bandwidth) or receiver device (e.g. resolution). However, this papers concerns a particular use case of transcoding, in which the coded bitstream is transcoded in order to replace an area of video frames, which is relatively small with respect to the video resolution. Some examples are 1) logo replacement/removal/enhancement, 2) hardcoded closed caption replacement (e.g. language change), 3) personalized advertisement and 4) insertion of sign language interpreter. In all these use cases, an initial source content is coded at the sender side once and is to be modified later on the delivery chain. 

The most straightforward approach to addressing the aforementioned scenarios is to perform a full decode-modify-encode operation. In this process, the compressed video from the source is decoded and decompressed to obtain the raw decoded pixel representation. Subsequently, the specific area to be replaced is identified, and the new content is inserted in its place. Finally, the modified signal is re-encoded to generate a new compressed video stream that incorporates the desired modifications.

There are two primary drawbacks associated with any solution that involves re-encoding a significant portion of the frames. Firstly, re-encoding a large area of frames requires significant computational resources and time, which can be inefficient in terms of both energy consumption and the cost of video delivery ecosystem deployment. Secondly, repeated re-encoding of frames can result in quality degradation. Each encoding process introduces compression artifacts and information loss, which can accumulate and adversely affect the overall visual quality of the video. This degradation is particularly noticeable when the replacement area is relatively small compared to the entire frame, as the re-encoding process affects a larger portion of the original content. %Preliminary experiments with one re-encoding pass show that this aspect can cause a performance drop ranging from negligible to approximately 6\% in terms of BD-BR \cite{bjontegaard2001calculation}. When using different codec implementations (\textit{e.g.}, x265, HM), the drop became more significant due to reduced reuse of coding decisions from the input bitstream

% The most intuitive solution to above scenarios is to carry out a full decode-modify-encode operation. To this end, the source compressed video is decoded and decompressed to be represented with raw decoded pixels. Then a replacement area is identified and consequently replaced by the new content. Finally, the modified signal is re-encoded to give a new compressed video stream with the required modifications. % This solution has three main drawbacks. First, it degrades the compression efficiency of the end-user displayed video, since, as it is widely known, multiple-compression causes damage to the quality to bitrate trad-eoff. Second, the re-encoding process could be costly (usually performed on public cloud servers). And third, the re-encoding process can cause additional latency in live video communication scenarios.

% \begin{figure}
%     \centering
%     \includegraphics[width=0.48\textwidth, angle=0]{figs/multi-encode.pdf}
%     \caption{Encoding a source $X$ and transcoding it multiple times ($T_i$, $i$=$1,..,N$) on a delivery chain, before reaching user, where $E$ and $D$ are the encoder and the decoder, respectively.}
%     \label{fig:full-transcode}
% \end{figure}

Another solution that is sometimes employed to tackle certain aspects of the aforementioned problem is to define independently coded areas within the video signal, specifically covering the replaceable area \cite{hannuksela2004isolated}. In this approach, when a replacement is required, transcoding is only applied to the target areas of the coded bitstream, avoiding the need for a full decode-modify-encode operation. 
% Another solution that sometime is used to address some forms of the above problem is to define independently coded areas on video signal covering the replaceable area \cite{hannuksela2004isolated}. Then upon request for the replacement, the transcoding is only applied on the target areas of the coded bitstream.
% Motion-Constrained Tiles Set (MCTS) in H.265/HEVC \cite{hevc,misra2013overview,mctspatent} standard and Sub-Pictures in H.266/VVC \cite{vvc,hamidouche2022versatile,vvchls} can be used for this purpose. Both MCTS and sub-picture techniques allow encoding with constraining motion vectors such that each piece (i.e. tile or subpicture depending on the standard) can be decoded and transmitted independently. Therefore, defining pieces borders such that replacement area falls into one piece, while constraining other pieces such that they do not refer to that piece, would address the above problem. 
Motion-Constrained Tiles Set (MCTS) in H.265/HEVC \cite{hevc,misra2013overview,mctspatent} and sub-Pictures in H.266/VVC \cite{vvc,hamidouche2022versatile,vvchls} are techniques that can be utilized to address the problem. Both methods allow encoding with constrained motion vectors, enabling independent decoding and transmission of specific pieces (tiles or sub-pictures). By defining borders to encompass the replacement area within a single piece while constraining others, the issue can be somehow resolved. 
% Although this solution is beneficial for certain use-cases (e.g. 360 video coding), it suffers from several drawbacks for the content replacement problem that this paper is attempting to address. In particular, if the replacement area is relatively small, then applying either MCTS or sub-pictures would require either very small pieces or large pieces with a small portion of replacement content. In the first case (i.e. small pieces), the total number of pieces is increased which could potentially be problematic as this parameter is usually chose carefully with respect to several other aspects (parallelism capacity, latency, etc.). In the second case (large pieces), as the replacement content occupies relatively a small portion of its piece, the re-encoding drawbacks (explained above) would still apply on a significant portion of the image. 
Although MCTS and sub-pictures can be beneficial for certain use cases, such as 360$^\circ$ video coding \cite{hannuksela2023vvc}, they may not be the most suitable solutions for addressing the content replacement problem described in this paper, when dealing with relatively small replacement areas. One drawback arises when using small pieces (\textit{i.e.} tiles or sup-pictures), as it increases the overall number of pieces, potentially causing issues related to parallelism capacity and latency. On the other hand, employing large pieces with a small portion dedicated to replacement content still results in significant portions of the image being subject to the re-encoding drawbacks mentioned earlier. %Therefore, while MCTS and sub-pictures have their advantages, they may not offer the optimal approach for efficiently addressing content replacement in scenarios involving smaller replacement areas.

% This papers proposes a method that re-encodes only a part of each frame that is collocated on the replace area and leaves other parts unchanged. If applicable depending on the content replacement, two advantages are offered by the proposed method. First, it could improve the compression efficiency by avoiding multiple compression on areas outside the replacement area. Second, it can significantly accelerate the transcoding process as it actually the time-consuming Rate-Distortion Optimization (RDO) only on the the replacement area and keeps using the initial syntax elements and coding decisions on other areas. 
This paper proposes a method that selectively re-encodes only the replacement area of each frame while leaving other parts unchanged. It offers two advantages: improved compression efficiency by avoiding redundant compression outside the replacement area, and faster transcoding by focusing on the time-consuming RDO only for the replacement area.
This paper is structured as follows. In Section~\ref{sec:method}, the proposed method is described. Section~\ref{sec:results} presents the testing conditions and the results obtained from the experiments. Finally, in Section~\ref{sec:conclusion}, the paper concludes with discussions on the limitations of the proposed method and potential directions for future research.

\section{Proposed partial transcoding}
\label{sec:method}
\begin{figure}
    \centering
    \includegraphics[width=0.48\textwidth, angle=0]{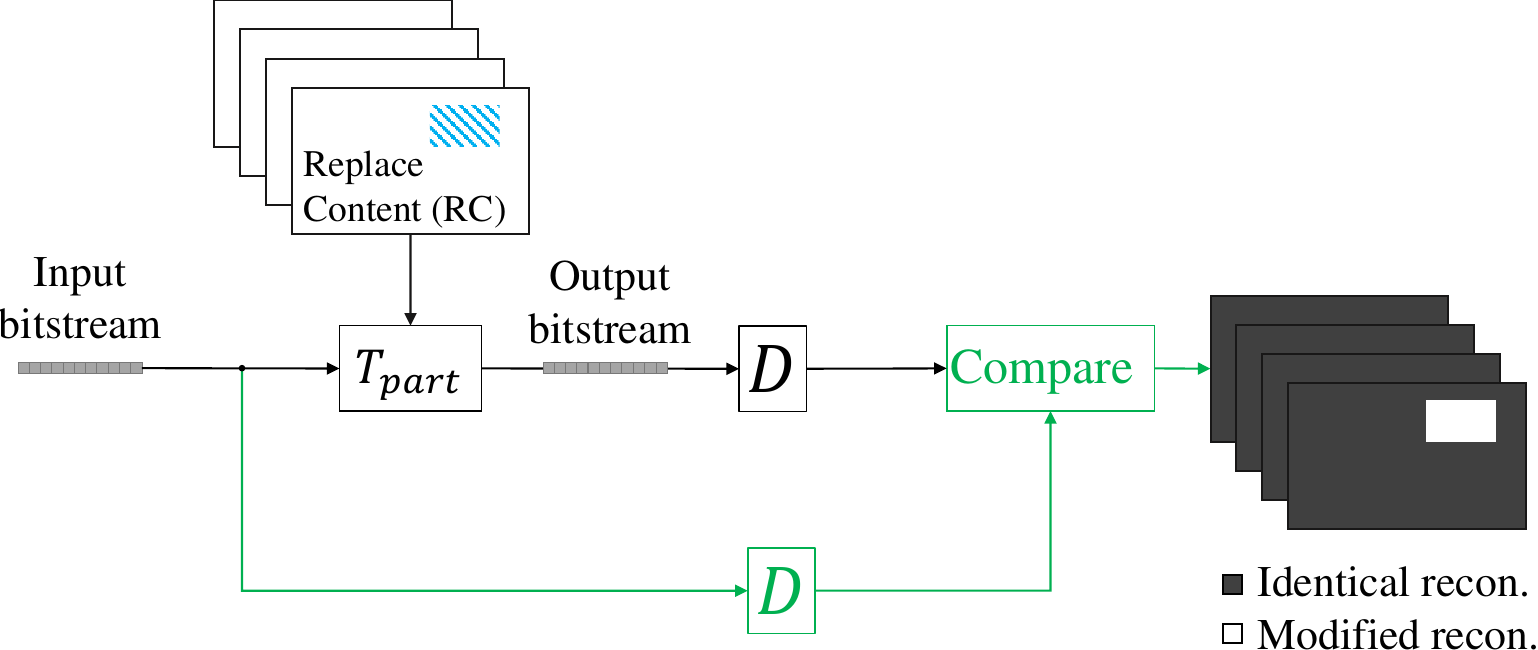}
    \caption{High level overview of the proposed partial transcoding, given a Replace Content (RC). The green path is informative to emphasize the identical reconstruction outside the Replace Area (RA).}
    \label{fig:diagram}
\end{figure}

Fig.~\ref{fig:diagram} provides a high-level overview of the proposed partial transcoding solution. The input to the partial transcoding module, labeled as $T_{part}$, consists of a coded bitstream and a Replace Content (RC), covering a Replace Area (RA). The module processes these inputs to generate an output coded bitstream. Notably, when the output bitstream is decoded by $D$, it differs from the input bitstream solely in the pixels specified by RC. 

% Fig. \ref{fig:diagram} depicts a high level view of the proposed partial transcoding solution. A coded bitstream and a Replace Content (RC) are given as input to a partial transcoding module -- denoted as $T_{part}$ -- to produce an output coded bitstream. The key feature of $T_{part}$ is that, when decoded by \textit{D}, the output bitstream is different from the input bistream only in the pixels indicated by RC. 

% The proposed method in this section is implemented for All Intra (AI) coding configuration. However, inter coding configurations, including P and B slices require further attention that will be briefly reviewed in Section~\ref{sec:conclusion}.

% The proposed process of producing an output bitstream is carried out at the frame-level. This consists of four main steps, namely 1) decoding and buffering the input bitstream, 2) CU categorization, 3) partial re-encoding of collocated CUs, and finally 4) re-encoding of the CUs impacted by the replace area. These steps are described in the following sections.

The process of generating the output bitstream in the proposed method is performed at the frame-level and involves four main steps. Firstly, the input bitstream is decoded and buffered. Secondly, a Coding Unit (CU) categorization step is applied. Next, partial re-encoding is performed on CUs collocated with the RA. Finally, the re-encoded CUs are re-written in the output bitstream. %These steps will be further detailed in the subsequent sections.

\subsection{Decoding and buffering the input bitstream}
A normal decoding process typically involves two stages: parsing syntax elements and reconstructing CUs. In the first step of the proposed method, both stages are executed as usual, resulting in two intermediate buffers: the syntax elements buffer and the reconstructed pixels buffer. The syntax elements buffer consists of pairs of $<$\textit{bin}, \textit{ctx}$>$, where \textit{bin} represents the binarized element that undergoes entropy coding using the context model defined by \textit{ctx}. The elements in this buffer are ordered based on their normative coding order. On the other hand, the reconstruction buffer simply contains the decoded pixels of the frame. 

% Let the decoding process consist of two stages of ``parsing'' syntax elements by the entropy decoder and ``reconstruction'' of CUs by applying prediction and residual decoding, defined by the parsed syntax elements. The first step of the proposed method carries out both stages normally and produces two intermediate buffers, namely syntax elements buffer and reconstructed pixels buffer. The syntax elements buffer is formed of pairs of $<$\textit{bin}, \textit{ctx}$>$, where \textit{bin} is the binarized element whose entropy coding is carried out using the method defined with \textit{ctx}. Pair elements in this buffer are ordered with their normative coding order. The reconstruction buffer, however, simply contains the decoded pixels of the frame.

\subsection{CU categorisation}
% Two types of spatial dependencies, namely intra prediction mode and loop filtering, are taken into account for identifying the last line of CUs implicated in the RA to avoid error propagation due to the content replacement. In particular, let $L$ be the maximum number of lines of pixels from a current CU that can be involved in reconstruction of a next CU. In HEVC, application of intra prediction mode involves one line of pixels from a neighboring CU, while the Deblocking filter involves up to three pixels. Therefore, $L=3$ is considered.

\begin{figure}
    \centering
    \includegraphics[width=0.48\textwidth, angle=0]{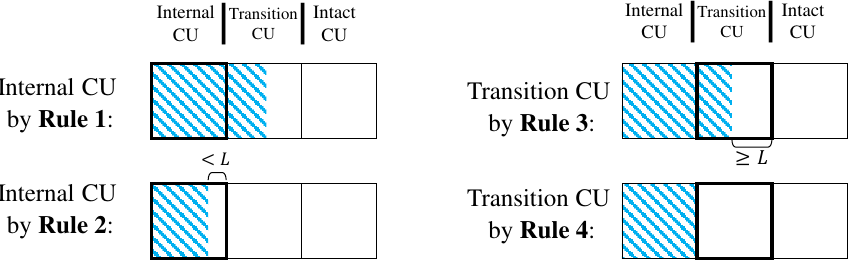}
    \caption{CU categorization using the defined rules, where the blue parts indicate the Replace Area (RA)}
    \label{fig:partial}
\end{figure}

% If part of the content of a given CU is to be replaced, then it is imperative to avoid establishing any spatial dependency to those part from other CUs that are not impacted by the replacement process. To do so, the coordinates of the Replace Area (RA) are used and CUs are categorized into three groups, namely 1) \textit{internal CUs}, 2) \textit{transition CUs} and 3) \textit{intact CUs}. Following rules are used for this categorization:
When replacing a portion of the content within a CU, it is crucial to prevent any spatial dependency between the replaced area and unaffected areas of other CUs. To achieve this, the coordinates of the RA are utilized to categorize CUs into three groups: 1) internal CUs, 2) transition CUs, and 3) intact CUs. The following rules, also visualized in Fig.~\ref{fig:partial}, are employed for this categorization:

\begin{itemize}    
    \item \textbf{Rule 1}: If all pixels of the CU are within the RA, then the CU is an \textit{internal CU}.
    \item \textbf{Rule 2}: If some pixels of the CU are located outside the RA and they represent less than $L$ lines of pixels, whether vertical or horizontal, then the CU is an \textit{internal CU}.
    \item \textbf{Rule 3}: If some pixels of the CU are located outside the RA but they represent more than or equal to $L$ lines of pixels, whether vertical or horizontal, then the CU is a \textit{transition CU}.
    \item \textbf{Rule 4}: If no pixel from the CU is located within the RA and the CU has spatial dependency to an \textit{internal CU}, then current CU is a \textit{transition CU}. 
    \item \textbf{Rule 5}: If none of the above cases are true, then the CU is an \textit{intact CU}.
\end{itemize}

The proposed method takes into account two spatial dependencies, intra prediction mode, and loop filtering, to identify the last line of CUs affected by the content replacement and prevent error propagation. In HEVC, the application of intra prediction mode involves one line of pixels from a neighboring CU, while the deblocking filter can impact up to three pixels. Therefore, to address these dependencies, a maximum number of lines, $L=3$, is considered, ensuring that the identified last line of CUs implicated in the RA accounts for both intra prediction and loop filtering effects.

\subsection{CU re-encoding}
% Depending on category, CUs might be re-encoded by going through the Rate-Distortion Optimization search to make new coding decisions, or re-written by using the same coding decisions as in the input coded bitstream. In particular, internal and transition blocks are re-encoded, while intact blocks are re-written. The process of re-encoding also depends on the category of the CUs. In nutshell, internal replace CUs are encoded arbitrarily, while the transition replace CUs have to be re-encoded using a lossless coding mode with the same partitioning decision. 
Depending on the category of CUs, they undergo different treatments during transcoding. Internal and transition CUs are re-encoded using the Rate-Distortion Optimization (RDO) search, while intact CUs are re-written using the original coding decisions. The re-encoding process depends on the CU category: internal replace CUs are encoded arbitrarily, while transition replace CUs are re-encoded using a lossless coding mode with the same partitioning decision.

\subsubsection{Lossy re-encoding of internal CUs}
To optimize compression efficiency, the internal replaced CUs are encoded in a lossy mode. For each CU, the target content is retrieved from the RC signal and encoded using the existing RDO process of the underlying codec. As a result, the pixels within the internal CUs can have various combinations of coding modes and partitioning, allowing for flexibility in the encoding process. 
% To maximize the compression efficiency, the internal replace CUs are encoded in lossy mode. For each CU, first the target content is fetched from the RC signal, then it is arbitrarily encoded using the typical Rate-Distortion Optimization (RDO) of the underlying codec. That being said, the pixels of the internal CUs are allowed to have any arbitrary combination of coding mode and partitioning.

\subsubsection{Lossless re-encoding of transitions CUs} 
% Transition replace CUs are losslessly coded such that pixels that are used in any form of spatial coding dependency would be reconstructed similarly from the output bitstream. To this end, the pixels of each transition replace CU are first categorized into reference and non-reference pixels, depending on the above dependencies. Non-reference pixels are filled with the RC, while reference pixels are filled using their collocated pixels from the reconstruction buffer. 
Transition replaced CUs are subject to lossless coding to ensure consistent reconstruction of pixels involved in any spatial coding dependency. In this process, the pixels within each transition CU are categorized into reference and non-reference pixels. Non-reference pixels are replaced with the content from the RC signal, while reference pixels are filled using corresponding pixels from the reconstruction buffer. This approach guarantees that the reconstructed reference pixels maintain their original values from the input bitstream, allowing for consistent spatial coding dependencies within the transition replace CUs.

\subsection{Bitstream re-writing}
\begin{figure}[h]
    \centering
    \includegraphics[width=0.48\textwidth, angle=0]{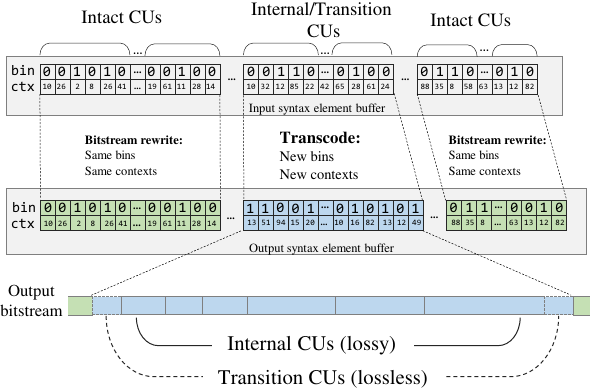}
    \caption{An example of computing output bitstream from an input syntax element buffer, by partial re-encoding of internal/transition replace CUs.}
    \label{fig:rewrite}
\end{figure}

% Regardless of the above choice for a given CU, the output of the re-encoding process is a new set of syntax elements. This set replaces the existing syntax elements for the CU and partially forms the output syntax elements buffer. On the contrary, the syntax elements of intact CUs are identically copied in the output syntax element buffer. By doing so, the output buffer will perfectly respect the specification of the underlying standard (\textit{e.g.} HEVC) and is able to be parsed and decoded. The process of forming the output buffer is depicted in Fig.~\ref{fig:rewrite}, where buffered syntax elements of replaced CUs are modified (blue), while intact CUs keep their initial syntax elements (green).
Irrespective of the aforementioned choice for a given CU, the re-encoding process generates a new set of syntax elements as the output. These new syntax elements replace the existing ones for the CU, contributing to the output syntax elements buffer. Conversely, intact CUs have their syntax elements directly copied into the output buffer. This approach ensures compliance with the specifications of the underlying standard, such as HEVC, allowing for proper parsing and decoding. Fig.~\ref{fig:rewrite} illustrates the process of forming the output syntax element buffer, where the syntax elements of replaced CUs are modified (blue), while intact CUs retain their initial syntax elements (green).

% Once the output syntax element buffer of the whole frame is prepared, it is written into the output bitstream using the entropy encoder. It is important to note that, even though intact CUs keep their intial syntax elements, if they are placed after the first replace CU in the scan order, then their entropy encoding (e.g. CABAC) would result in new states of context models. This is due to the fact that by replacing a part of the past symbols, probability models that are computed based them would also change. 
The prepared output syntax element buffer for the entire frame is subsequently encoded into the output bitstream using the entropy encoder. It is worth noting that although intact CUs retain their initial syntax elements, their entropy encoding (\textit{i.e.}, CABAC) can still lead to new context model states if they appear after the first replace CU in the scan order. This is because replacing a portion of past symbols affects the probability models computed based on them, resulting in changes to the context models used for entropy encoding.

\begin{figure}[h]
    \centering
    \begin{tabular}{cc}
    \includegraphics[width=0.20\textwidth]{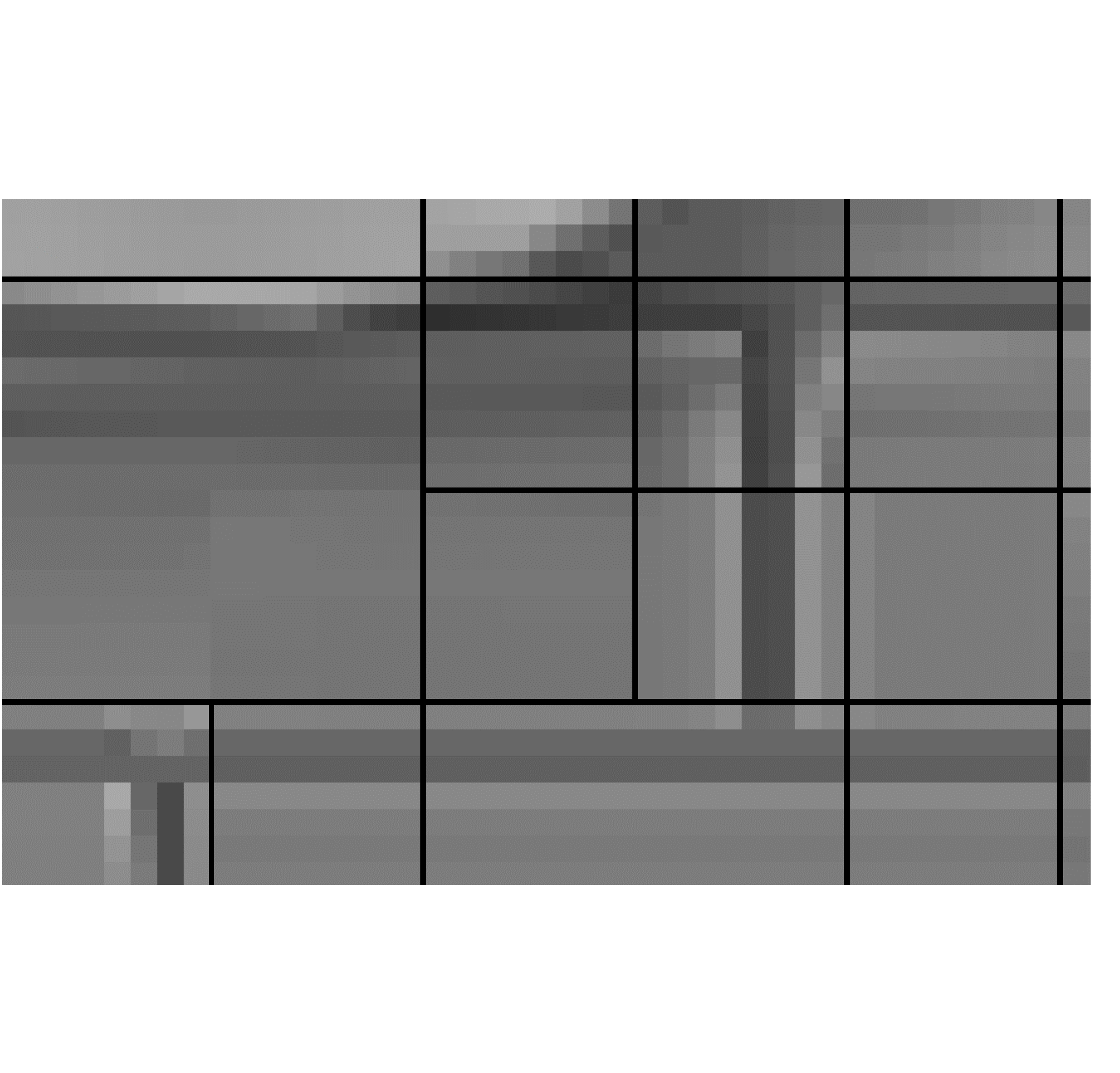}
    &
    \includegraphics[width=0.20\textwidth]{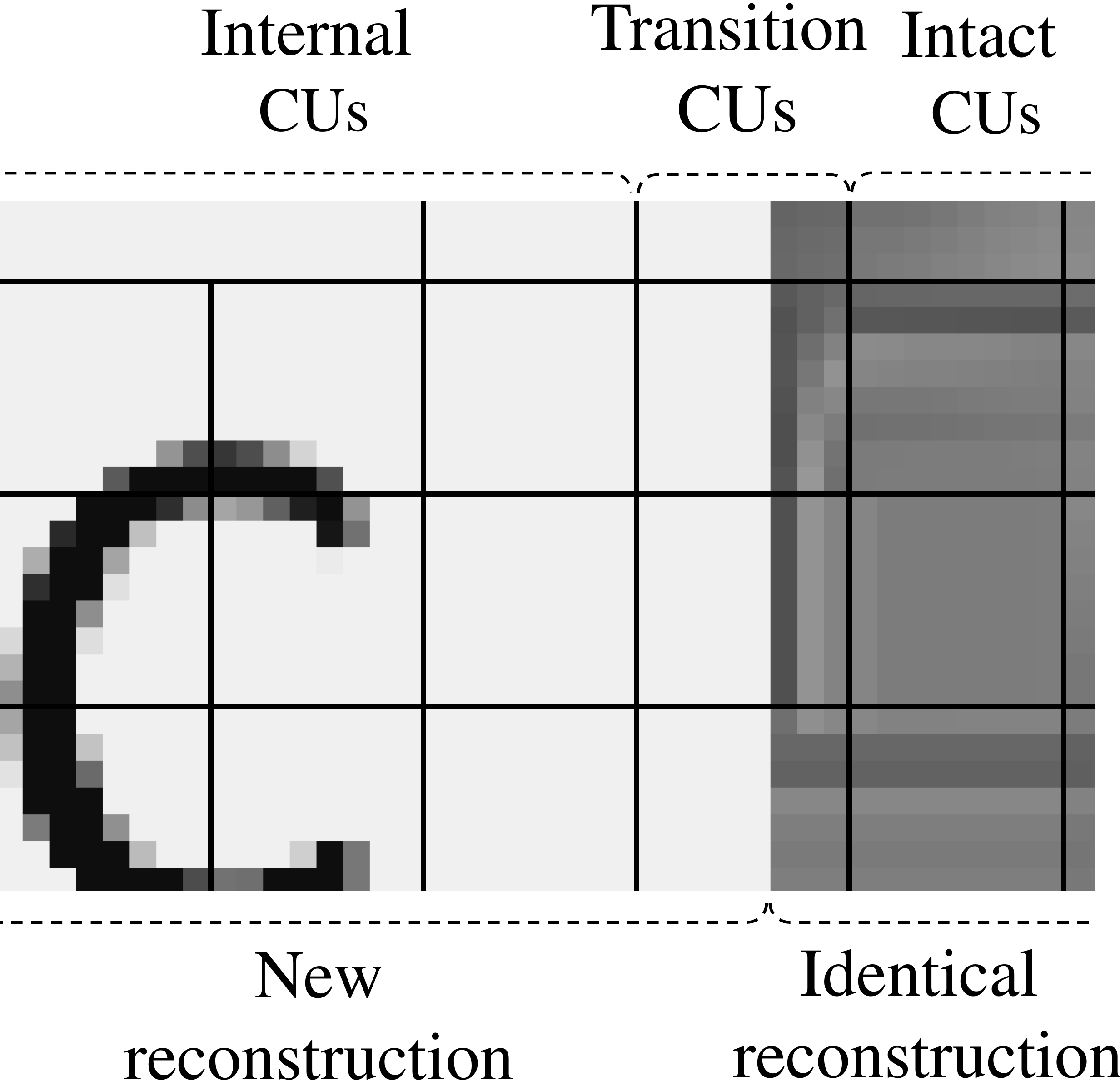}
    \\ 
    \small{Decoded input bitstream}  &
    \small{Decoded output bitstream}

    \end{tabular}    
    \caption{Reconstruction and partitioning of different CUs after the content replacement by the proposed partial transcoding.}
    \label{fig:reconstruction}
\end{figure}

% Fig.~\ref{fig:reconstruction} shows an actual example of the implementation of the above process. In this figure, the left part of the input bitstream is replaced by a by a new content (insertion of the character 'C' on a plain background) and is partially transcoded. As can be seen, the right side pixels of the transition CUs as well as all pixels of the intract CUs are identically reconstructed in the output bitstream, while the replace content is freely (new partitioning) coded.
Fig.~\ref{fig:reconstruction} illustrates an actual implementation example of the aforementioned process. The input bitstream's left portion is replaced with new content, specifically the insertion of the character `C' on a plain background, and undergoes partial transcoding. The output bitstream demonstrates that the right-side pixels of the transition CUs and all pixels of the intact CUs are reconstructed identically. Meanwhile, the replacement content is freely encoded with new partitioning decisions, providing flexibility in the coding process.

% =========================================================
% =========================================================
% =========================================================
% =========================================================
% =========================================================
% =========================================================
% =========================================================
% =========================================================
% =========================================================

\section{Experimental results}
\label{sec:results}

\subsection{Test conditions}
% The proposed method was implemented on top of the HEVC test Model (HM) version 17. Correspondingly, in all experiments is the full-transcode solution using the HM. Moreover, the test sequences in the JVET Common Test Condition \cite{ctc} were used for encoding and transcoding. Two main metrics have been used for the performance evaluation. First metric is BD-BR \cite{bjontegaard2001calculation}, which represents the average of bitrate saving in the same level of quality. Aligned with the CTC, the encodings have been carried out int fixed QP with four values of 22, 27, 32, 37. To compare the transcoding acceleration of the proposed method, the second metric is chosen to be the transcoding complexity, which is measured in terms of encoding time. 
The proposed method was implemented\footnote{A limited prototype of the modified HM encoder with the essential elements of the proposed method has been made available here: \href{https://github.com/b-com/HevcPartialTranscoding}{https://github.com/b-com/HevcPartialTranscoding}} using HEVC Test Model (HM) v17.0. The anchor method applies a full decode-replace-encode scheme, where both encoding trials (\textit{i.e.} source encoding and the encoding after the content replacement) are carried out with the All Intra (AI) configuration of the HM encoder. The JVET Common Test Conditions (CTC) test sequences were utilized for encoding and transcoding \cite{ctc}. Two main metrics were used for performance evaluation: BD-BR \cite{bjontegaard2001calculation} and transcoding complexity. BD-BR measures the average bitrate saving at the same quality level, while transcoding complexity is measured in terms of time saving. The encodings were performed at fixed QP values (22, 27, 32, 37) in line with the CTC. 

\begin{table*}
\caption{BD-BR and time saving (TS) performance of the proposed method (with respect to HM-17.0) in different settings of parameters, namely the occupation rate and the replace position. Negative and positive percentages mean gain and loss in terms of bitrate saving, respectively.}
\label{tab:bdr}
\centering
\def\arraystretch{1.2}
\begin{tabular}{c|ccccc|ccccc|ccccc}
%\begin{tabular}{C{0.03\textwidth}P{0.05\textwidth}P{0.04\textwidth}P{0.0001\textwidth}P{0.05\textwidth}P{0.04\textwidth}P{0.0001\textwidth}P{0.05\textwidth}P{0.04\textwidth}}
%\begin{tabular}{C{0.03\textwidth}|P{0.03\textwidth}P{0.03\textwidth}P{0.03\textwidth}P{0.03\textwidth}P{0.03\textwidth}|P{0.03\textwidth}P{0.03\textwidth}P{0.03\textwidth}P{0.03\textwidth}P{0.03\textwidth}|P{0.03\textwidth}P{0.03\textwidth}P{0.03\textwidth}P{0.03\textwidth}P{0.03\textwidth}}
\hline
\hline
\multirow{2}{*}{\textbf{Class}} 	& \multicolumn{5}{c|}{Occupation: 1\%} & \multicolumn{5}{c|}{Occupation: 10\%} & \multicolumn{5}{c}{Occupation: 50\%} \\
\cline{2-16}	
									& TL  		& TR  			& BR 			& BL 		& C 			& TL  		& TR  		& BR  		& BL 		& C 		& TL  		& TR  		& BR  		& BL 		& C \\
\hline	
\textbf{A}                      	&   -4.1\%  &	 -3.3\%  	& 	-2.2\%  	& -2.7\%  	& -3.1\%    	&  -4.7\%	&   -4.4\%	&   -3.6\%	&   -3.7\%	&   -4.2\%	&  +3.0\%	&   +4.1\%	&   +4.4\%	&   +3.2\%	&   +4.6\%   \\
\textbf{B}                      	&   -3.7\%  &	 -3.0\%  	& 	-2.2\%  	& -3.0\%  	& -3.5\%    	&  -4.5\%	&   -3.6\%	&   -3.2\%	&   -3.4\%	&   -3.8\%	&  +4.0\%	&   +4.8\%	&   +3.8\%	&   +3.8\%	&   +4.1\%   \\
\textbf{C}                      	&   -3.5\%  &	 -3.2\%  	& 	-2.0\%  	& -2.9\%  	& -3.3\%    	&  -3.9\%	&   -4.3\%	&   -3.1\%	&   -3.7\%	&   -3.9\%	&  +2.4\%	&   +4.3\%	&   +3.5\%	&   +4.4\%	&   +4.8\%   \\
\textbf{D}                      	&   -3.9\%  &	 -3.2\%  	& 	-2.3\%  	& -2.8\%  	& -3.0\%    	&  -4.2\%	&   -4.2\%	&   -3.5\%	&   -4.2\%	&   -3.8\%	&  +3.3\%	&   +3.5\%	&   +3.8\%	&   +3.4\%	&   +4.5\%   \\
\textbf{E}                      	&   -3.9\%  &	 -2.9\%  	& 	-2.1\%  	& -3.0\%  	& -3.5\%  		&  -4.5\%	&   -3.7\%	&   -2.9\%	&   -3.5\%	&   -4.4\%	&  +2.8\%	&   +4.3\%	&   +4.8\%	&   +4.5\%	&   +4.0\%   \\
\hline							
\multirow{2}{*}{\textbf{Average}}	&  	-3.8\%  &  	 -3.1\%  	&  -2.1\%		& -2.8\%  	& -3.3\%  		&  -4.6\%  	&  -4.0\%   &  	-3.5\% 	& 	-3.7\%   & 	-4.0\%  &  +3.1\%	&  +4.2\% 	&	+4.3\%	&  +3.4\%	&  +4.3\%   \\
\cline{2-16}
									& \multicolumn{5}{c|}{-3.02\%}    & \multicolumn{5}{c|}{-3.96\%}     & \multicolumn{5}{c}{+3.88\%}     \\
\hline	
\textbf{TS}                     	& \multicolumn{5}{c|}{83\%}    & \multicolumn{5}{c|}{65\%}     & \multicolumn{5}{c}{12\%}     \\
\hline
\hline
\end{tabular}

\end{table*}

\subsection{Parameters}
% Different parameters might impact the performance, including number of transcoding passes, Replace content, its relative size placement. To vary the replace content, the replace content signals were randomly selected from other test sequences. Moreover, the replace size and its placement have also been varied. In particular, three occupation rates of 1\%, 10\% and 50\% are considered that represent the ratio between the number of pixels in the replace area and the main content. For all values, the same aspect ratio as the main signal has been used for the replace signal. Similarly, five replace positions are considered, namely the four corner of the image Top-Left (TL), Top-Right (TR), Bottom-Right (BR), Bottom-Left (BL) and the center (C). However, due to space limitation, only one transcoding pass has been tested.
Various parameters can influence the performance of the proposed method, including the number of transcoding passes, the RC characteristics (\textit{e.g.} content complexity, relative size and placement). To introduce variation in the replace content, its signals were randomly selected from other test sequences in the CTC. Additionally, the replace size and its placement were diversified. Specifically, three occupation rates of 1\%, 10\%, and 50\% were considered, representing the ratio between the number of pixels in the RA and the main content. The replace signal maintained the same aspect ratio as the main signal. Furthermore, five replace positions were examined: Top-Left (TL), Top-Right (TR), Bottom-Right (BR), Bottom-Left (BL), and the center (C) of the image. Due to space limitations, only one transcoding pass was tested.

\subsection{Performance}
% Table~\ref{tab:bdr} shows the performance of the proposed method. In terms of BD-BR, the proposed method offers -3.02\% and -3.98\% gain for occupation rates of 1\% and 10\%, respectively, while it results in +3.88\% loss when the occupation rate reaches 50\%. In terms of time saving, the method saves 83\%, 65\% and 12\% of the transcoding time for the occupation rates of 1\%, 10\% and 50\%, respectively.
Table~\ref{tab:bdr} shows the performance of the proposed method. BD-BR gains of -3.02\% and -3.96\% are achieved for occupation rates of 1\% and 10\% respectively, while a loss of +3.88\% is observed at 50\% occupation rate. Regarding time savings, the method reduces transcoding time by 83\%, 65\%, and 12\% for occupation rates of 1\%, 10\%, and 50\% respectively.

When the occupation ratio is high, the proposed method does not provide any BD-BR performance advantage, and the acceleration rate is negligible. This can be attributed to two factors. Firstly, the high presence of lossless blocks at the RA borders introduces a significant rate overhead, leading to a reduction in bitrate savings. Secondly, since a large portion of the frame needs to be re-encoded, the overall transcoding process is not accelerated significantly. 

When the replacement position is closer to the Top-Left corner of the image, a slightly better BD-BR gain is observed. This can be attributed to the fact that the quality of the Top-Left reconstructed CUs has a greater impact on the neighboring CUs. As the encoding process progresses towards the Bottom-Right corner, the encoding decisions made in the early stages have a cascading effect on subsequent CUs. Therefore, any improvement in the quality of the reconstructed CUs in the Top-Left area has a more significant influence on the overall image quality, resulting in slightly higher compression gains.

\subsection{Other observations}
Important observations from additional experiments, which cannot be included here due to space constraints, are briefly presented. First, it has been assessed that the method tends to always increase rate (even when it provides a BD-BR gain) and its cause is believed to be the insertion of lossless coded blocks. This increase is less than 0.1\%, around 3\% and more than 10\% for occupation rates of 1\%, 10\% and 50\%, respectively.

A second observation was that the performance varies depending on bitrate. In particular, poorer BD-BR performance was observed in lower bitrates, which again is believed to be related to the rate overhead introduced by lossless blocks, resulting in reduced bitrate savings. However, in the middle bitrate range, the method shows high performance gains with a moderate level of acceleration. Finally, At high bitrates, the gains in terms of bitrate reduction become smaller, but there is a significant acceleration benefit. This is because the portion of the frame being re-encoded is reduced, resulting in notable time savings during the encoding process. Therefore, the method's performance varies with different bitrates, with a trade-off between bitrate savings, acceleration, and the impact of lossless blocks.

% =========================================================
% =========================================================
% =========================================================
% =========================================================
% =========================================================
% =========================================================
% =========================================================
% =========================================================
% =========================================================

\section{Conclusion and Perspectives}
\label{sec:conclusion}
% This paper presents a partial transcoding method that avoids a full scheme with the chain of decode-replace-encode operations. This method targets use-cases where the are of the partial content change is relatively small aims, at it aims at improving both transcoding time and compression efficiency. To to so, the proposed method identifies spatial dependencies between CUs on the border of the replace area and avoids a propagation of the decoding mismatch by losslessly coding the reference pixels. Although current method has proved efficienct in the tested settings, two main limitations are known to authors.
This paper introduces a partial transcoding method that eliminates the need for a complete decode-replace-encode chain, focusing on scenarios where the area of content modification is relatively small. The method aims to enhance both transcoding time and compression efficiency. It achieves this by identifying spatial dependencies between CUs at the border of the replace area and preventing the propagation of decoding mismatch through lossless coding of reference pixels. While the proposed method has demonstrated effectiveness in the tested settings, two main limitations are known to authors.

\subsection{Inter-frame coding}
% Current method performs is implemented only for I-frames, while in practical applications of transcoding, P and B frames are typically used. The current method, as such, is not capable of addressing intera frames, since it only takes into account spatial dependencies. However, our preliminary results show that with a few additional constraints a similar partial transcoding scheme could be applied also for inter frames. First, a motion constraint needs to be imposed at the encoder side, such that no pixel within the Replace Area (RA) is referred to by a motion compensated CU outside the RA. In other words, if the sender detects a potentially replaceable area (\textit{e.g.} logo, subtitle \textit{etc.}) in the source video, then its initial encoder must avoid motion compensation from areas around the replaceable area to pixels within it. Second, the RA has to be determined at the encoder side and somehow signaled to potential transcoders who might be interested to replace the RA. For instance, one might define new Supplementary Enhancement Information (SEI) messages to carry auxiliary masks of replaceable areas \cite{masksei}.
The current implementation of the method is limited to I-frames, while practical transcoding applications involve P and B frames. This limitation arises because the current method solely considers spatial dependencies and does not address inter frames. However, preliminary results indicate that with additional constraints, a similar partial transcoding scheme can be applied to inter frames. To achieve this, two requirements must be met. Firstly, a motion constraint should be imposed at the encoder side, ensuring that no pixel within the RA is referenced by a motion-compensated CU outside the RA. Secondly, the RA needs to be determined at the encoder side and signaled to potential transcoders interested in replacing the RA. One approach could involve defining new Supplementary Enhancement Information (SEI) messages to convey auxiliary masks of replaceable areas \cite{masksei}.

\subsection{Codecs other than HEVC}
% Proposed algorithm is particularly specified to be integrated in HEVC transcoders. However, one can extend its concept also to other codecs as long as they provide CU-levels tools to enable lossless coding. Our preliminary studies show that this requirement is met by Versatile Video Coding (VVC), as it allows CUs to be encoded in lossless using the Intra Block Differential Pulse-Code Modulation (BDPCM) tool \cite{bdpcm}. In VVC, some other adaptations are required. For instance, one should also take into account Block Vectors (BV) of the Intra Block Copy (IBC) tool that allows dependencies between CUs in the same block that are not necessarily adjacent \cite{ibc}. Moreover, the parameter $L$ -- described in Section II -- must be adapted as in VVC the deblocking filter is longer and also, Multiple Reference Line (MRL) allows further reference lines for intra prediction \cite{mrl}.
The proposed algorithm is primarily designed for integration into HEVC transcoders. However, the concept can be extended to other codecs that offer CU-level tools for lossless coding. Our preliminary studies indicate that this requirement is met by Versatile Video Coding (VVC), which provides the Intra Block Differential Pulse-Code Modulation (BDPCM) tool for lossless CU encoding \cite{bdpcm}. Adapting the algorithm for VVC would involve considering additional factors. For example, the Block Vectors (BV) of the Intra Block Copy (IBC) tool, which allows dependencies between non-adjacent CUs within the same block, need to be taken into account \cite{ibc}. Furthermore, the parameter $L$ described in Section II needs to be adjusted since VVC has a longer deblocking filter and Multiple Reference Line (MRL) allows additional reference lines for intra prediction \cite{mrl}. 
% \newpage
\balance
\bibliographystyle{IEEEbib}
\bibliography{myBib}

\end{document}